# [1]An Ultra-Specific Image Dataset for Automated Insect Identification

D.L.Abeywardhana[1,*], C.D.Dangalle[1], Anupiya Nugaliyadde[2,3], Yashas Mallawarachchi[3]

[1]*University of Colombo, Colombo, Sri Lanka*

[2]*Murdoch University, Perth, Australia*

[3]*Sri Lanka Institute of Information Technology, Malabe, Sri Lanka*

**Abstract**

Automated identification of insects is a tough task where many challenges like data limitation, imbalanced data count, and background noise needs to be overcome for better performance. This paper describes such an image dataset which consists of a limited, imbalanced number of images regarding six genera of subfamily Cicindelinae (tiger beetles) of order Coleoptera. The diversity of image collection is at a high level as the images were taken from different sources, angles and on different scales. Thus, the salient regions of the images have a large variation. Therefore, one of the main intentions in this process was to get an idea about the image dataset while comparing different unique patterns and features in images. The dataset was evaluated on different classification algorithms including deep learning models based on different approaches to provide a benchmark. The dynamic nature of the dataset poses a challenge to the image classification algorithms. However transfer learning models using softmax classifier performed well on current dataset. The tiger beetle classification can be challenging even to a trained human eye, therefore, this dataset opens a new avenue for the classification algorithms to develop, to identify features which human eyes have not identified.

# 1. Introduction

There are millions of species on earth, in which 1.2 million have already been formally described [35]. Some species have been identified and described using molecular techniques while morphology and morphometrics have been used to identify others. However, the application of molecular techniques requires considerable expertise knowledge, high cost and time while morphology-based identification poses challenges in identifying cryptic and less abundant taxa [5, 19]. Difficulties in morphological identification are mainly due to the number of visually similar categories which are required to identify inter-class similarity. Therefore, traditional species identification requires expert knowledge of each species in detail. Also, species identification consumes a large amount of time. The lack of expert knowledge, cost of species identification and high time consumption has shown importance towards image classification using Artificial Intelligence [32].

---

[1] This is a pre-print of the manuscript that's currently under review in Multimedia Tools and Applications Journal, Springer.

The rapid improvement in machine learning models especially convolutional neural networks (CNN) image classification has shown dramatic performances. These models are capable of identifying features which are visible and non-visible to the naked eye to classify objects [7, 26]. However these models require a large number of training data from each class to learn the features, that are organized to form a feature vector, an arbitrary length vector which collects all the properties that are useful in describing the object under analysis [50] to enhance the model for better performance [27, 41, 46]. As a result, Ultra-specific classifications such as insect species classification which have inter-genus similarities in morphology, require an extremely large amount of image data to gain high validation accuracy[20, 38, 54]. Therefore, this is a challenging task for automated models to extract features to identify these insect species who are with enormous inter-class similarities (Fig. 1).

This paper introduces an ultra-specific data set on tiger beetles, which is one of the first datasets that focus on entomological specific image data. The tiger beetle images classification requires expert knowledge to classify images. These images are taken from various image sources and various environments. Each image has different noise-levels, different size and a different zoom levels on the beetle. Therefore, to classify these images using computational models are challenging. General computational models are not capable of handling noise and variations of the images to classify. Therefore, the dataset would provide a good platform for new image analysis methods for real-world ultra-specific image classification, with highly different noise levels and variations. In this paper, benchmarks are also set to show the responses to each machine learning algorithm.

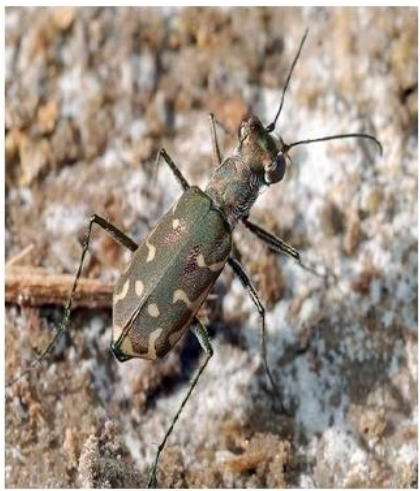 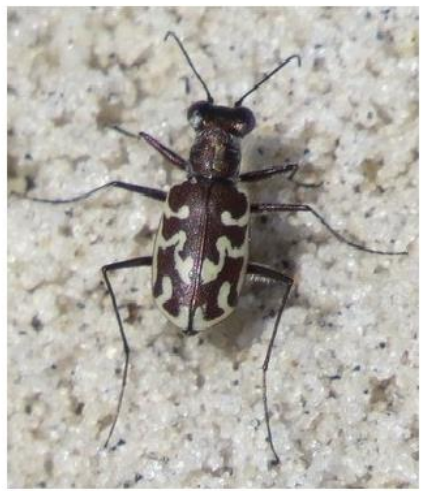

**(a)**        **(b)**

**Fig. 1.** Visually similar species from two different genera (a) *Myriochila distinguenda*
(b) *Calomera angulata*

Almost all the dominant image classification datasets used in computer vision tend to have a uniform (balanced) and copious distribution of images across object categories. As an example, the well-known, highly diverse ImageNet dataset[11] consist of more than 14 million images belonging to more than 20,000 categories with a typical category consisting of several hundred images. However, the feature vectors identified for such datasets are not substantially suitable for highly specific scenarios such as identification of zoological image sets of species belonging to the same family. To become more specific for animal species identification by machine learning, the Caltech-UCSD Birds-200-2011 Dataset [12] is comprised of bird images which has a 11,788 image collection and Stanford Dogs Dataset which contains images of 120 breeds of dogs from around the world, that has been built using images and annotation from ImageNet for the task of fine-grained image categorization [28]. Further, iNaturalist species classification and detection dataset, consisting 859,000 images from over 5,000 different species of plants and animals [22] and IP102, a large-scale dataset specifically constructed for insect pest recognition which contains more than 75,000 images belonging to 102 categories [52] have been developed. All above-mentioned datasets have at least more than 10,000 images in their collection, but this is often impractical, or even impossible, as in many real-world scenarios some species are more abundant and easier to photograph, to find resources and some are rare and endemic and not visible in the common environment frequently.

In some scenarios, the image collection in the databases has been taken in proper light condition, in specific dimensions (angle) with less background noise. Marques et al., [34] describes ant genera identification methodology using an image dataset as an online database on ant biology, the Antweb (http://www.antweb.org). The images in the above dataset were taken using powerful tools like Automontage and Leica microscope so that all the images in the dataset have same (high level) quality and also the dimensions of the images have been restricted to three views as frontal, lateral and dorsal. Larios et al., [30] proposes a methodology for the identification of stonefly larvae, an insect inhabiting in water. To capture high-quality images of stonefly larvae a mechanical apparatus has been buit where each specimen is manually inserted into acrylic and then pumped through a tube. During this process an infrared detector positioned along the tube detects the passage of the specimen and a side fluid jet captures the specimen using a QImaging MicroPublisher 5.0 RTV 5 megapixel colour digital camera attached to a Leica MZ9.5 high-performance stereo microscope at 0.63x magnification. Illumination has provided by gooseneck light guides powered by Volpi V-Lux 1000 cold light sources and diffusers have also been installed in this apparatus to reduce glare, specular reflections and hard shadows. Gutierrez et al.,[18] describes a pest detection and identification method on tomato plants. The dataset has been generated using both manually and automatically obtained images collected in a specific environment (Mendelu's cultivation chamber). In the manual procedure, all the images have been captured by AP-3200t-PGE and monochrome camera DataCam 2016R using a standard display system connected to a PC. Different types of lenses and lighting systems are also used when necessary. In the automated procedure, two GigE UI-5240CP cameras have been set-up and using Raspberry Pi 3 microcontroller both the camera and the movement structure are controlled. The microcontroller is programmed to take pictures both with and without artificial lighting in different directions and angles. However above set-ups can be impractical for some real-world problems when dealing with images taken in a normal environment deprived of proper light intensity with background noise. Furthermore, the rarity of the species prevents access to a large number of images for each class. Therefore, machine learning models struggle to achieve accurate classifications results. Hence it's proven that class imbalance, availability of limited images with inter-class similarity and availability of different quality images with huge background noise are properties of a real-world scenario where computer vision models should be able to deal with [22].

## 1.1 Research Contribution

The paper introduces a real-world tiger beetle image dataset which has been classified to genera level. This dataset uses images from various sources ranging from high-definition cameras to images scraped from the internet. The data set is created by an expert entomologist to test various machine learning models. Base-line machine learning models are used for experiments and are described as a benchmark for the dataset. Furthermore, this shows a path for future work required for a highly specified machine learning model for ultra-specific insect classification.

## 2. The Tiger Beetle Dataset [2]

The classification dataset of tiger beetles (Coleoptera, Carabidae, Cicindelinae) from Sri Lanka is comprised of a limited number of images. This dataset consists of images of tiger beetles belonging to six genera of tribes Cicindelini (ground-dwelling tiger beetles) and three genera of tribe Collyridini (arboreal tiger beetles) [1] (Fig. 2).

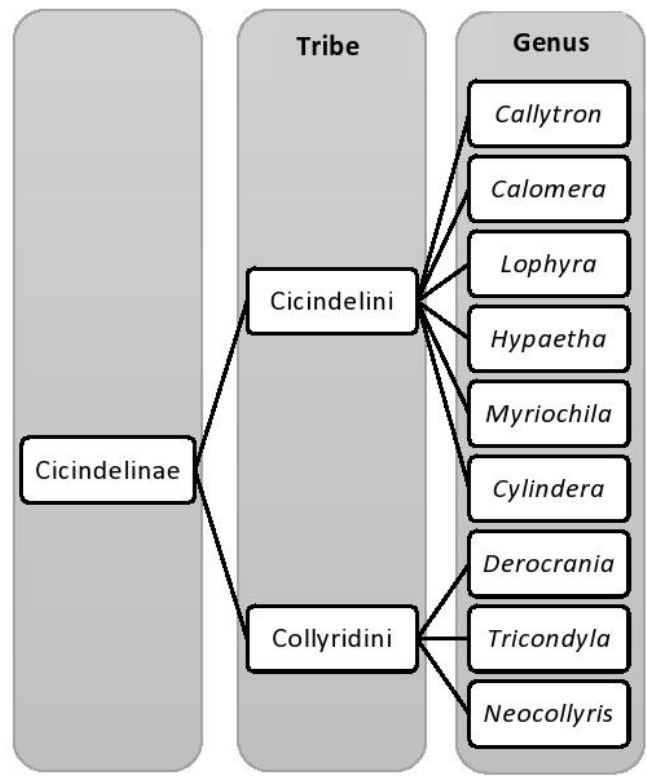

**Fig. 2.** Tiger beetle tribes and genera of the image dataset

There are significant morphological differences between species of tribe Cicindelini (Fig. 3a) and tribe Collyridini (Fig. 3b) to distinguish each tribe using visuals (Table 1).

Table 1. Morphological differences between tribe Cicindelini and tribe Collyridini

| Ground dwelling tiger beetles (Cicindelini) | Arboreal tiger beetles (Collyridini) |
|---|---|
| The body is not conical/ flask-shaped. | The body is conical/ flask-shaped. |
| Short pronotum. | Long, slender pronotum. |
| Elytra consist of different elytral patterns. | No elytral patterns. |
| Episterna of the metasternum not very narrow and not strongly furrowed. | Episterna of the metasternum very narrow and strongly furrowed. |
| Tarsal pads are present only in pro-thoracic legs of males. | Tarsal pads present in all legs of both sexes (*Collyris, Tricondyla*). |

Sources:[9, 14, 40]

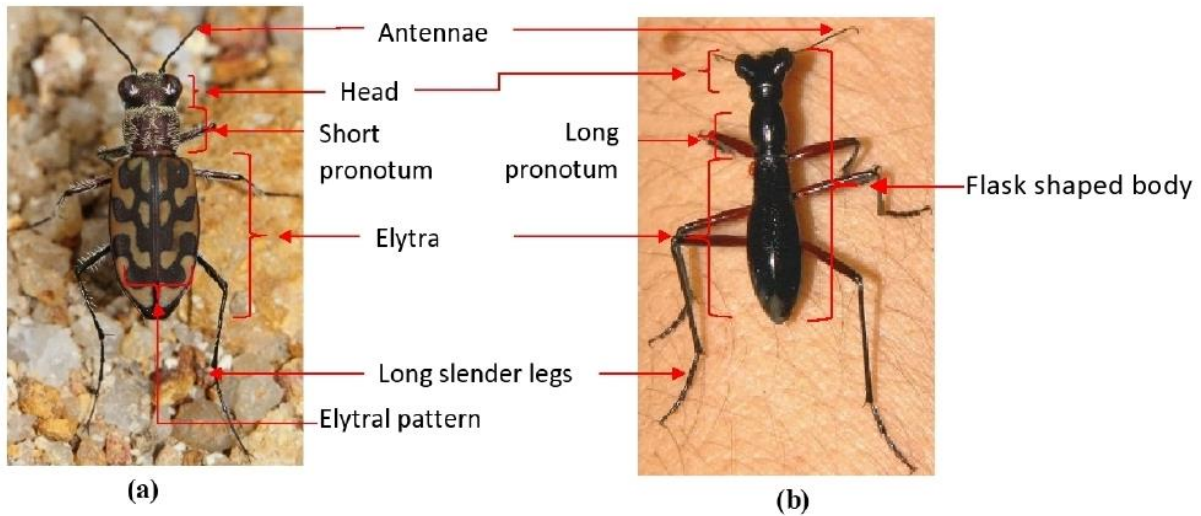

(a)                                      (b)

**Fig. 3.** (a) *Lophyra catena* of tribe Cicindelini, (b)  *Tricondyla* of tribe Collyridini

The different genera of tribe Cicindelini have distinctive elytral patterns that enable the identification of genera. Further, small variations of elytral patterns may also be seen in species within the same genera.

## 2.1 Tiger beetle genera used in the dataset

### 2.1.1 Tribe Cicindelini

#### I. Genus *Callytron*

The dataset of the present study used images of only *Callytron limosa* of genus *Callytron*. *Callytron limosa* has a unique elytral maculae pattern. Maculae have reduced to a narrow white continuous lateral band from the humeral angle to the apical spine (Fig. 4)

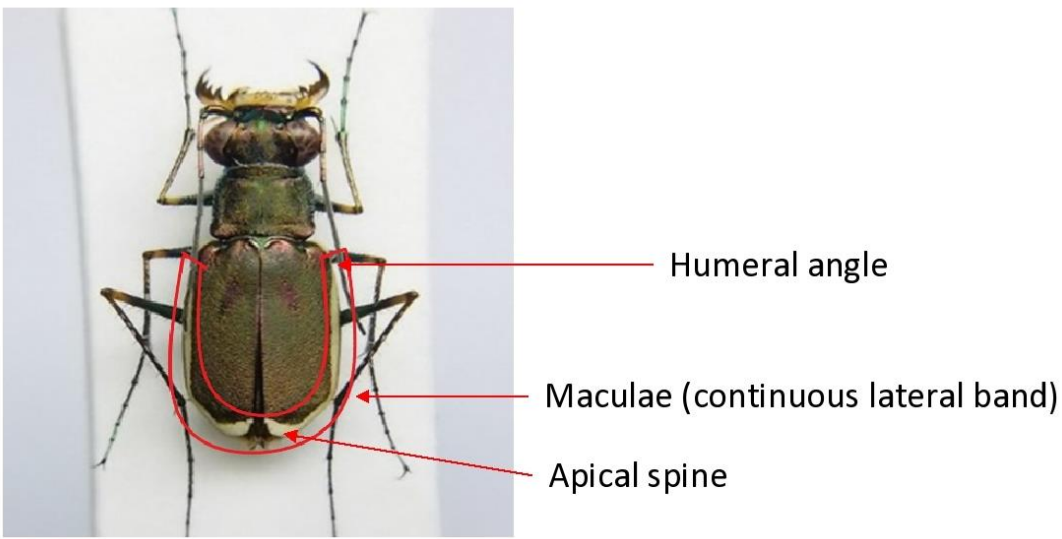

**Fig. 4.** *Callytron limosa*

#### II. Genus *Calomera*

The dataset utilized images of the species *Calomera angulata* and *Calomera cardoni* of the genus *Calomera*. Both species have a complete humeral lunule on elytra but can be differentiated using the middle band of elytra [3]. In *C. angulata* the middle band consists of a transverse portion that concaves anteriorly, while in *C. cardoni* the transverse portion of the middle band does not concave anteriorly. The terminal portion of the middle band of *C. angulata* is broadly connected while it is separate or narrowly connected in *C. cardoni* [10]. Further, *C. angulata* have a dark bronze coloured head and pronotum, with greenish punctures and white markings. White colour in elytra extends from the shoulders to the apex, with an interruption before the apical lunate patch (Fig. 5a).  The elytral surface of *C. cardoni* is dark brown or metallic brown with yellowish-white discontinuous lateral lunules. The middle band terminates medially to form a separate spot (Fig. 5b) [48].

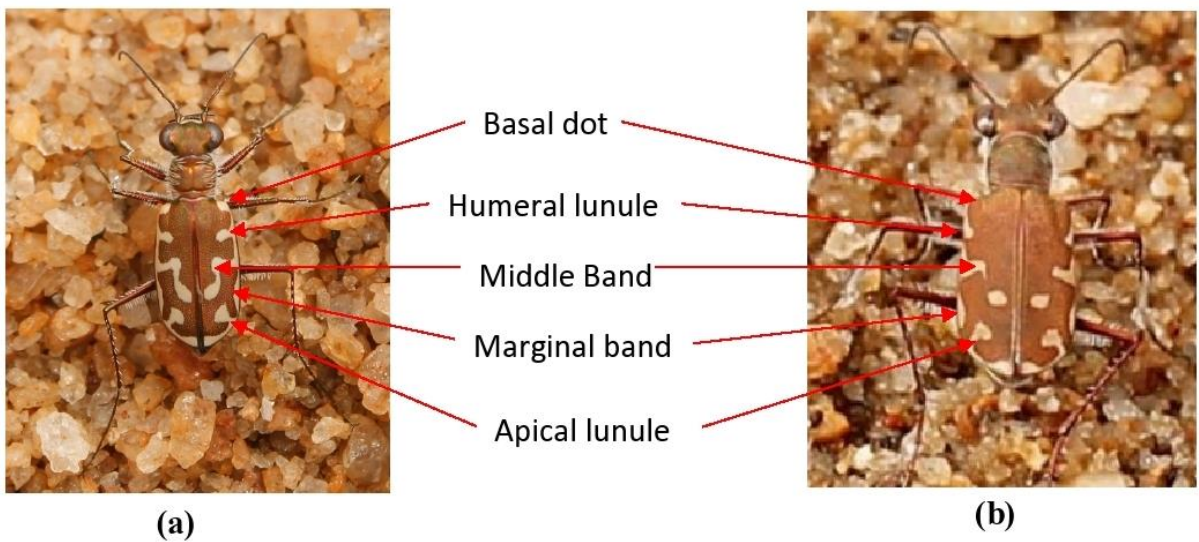

**(a)**                                                                                **(b)**

**Fig. 5.**  (a) *Calomera angulata; (*b) *Calomera cardoni*

## III.    Genus *Lophyra*

*Lophyra catena* and *Lophyra cancellata* of genus *Lophyra* are morphologically very similar. However, in *Lophyra catena* the genae of the head are setose, while in *Lophyra cancellata* they are glabrous (smooth and without hairs) [22]. Humeral lunule, the middle band and apical lunule of both the species of this genera are clearly visible and similar to each other. Above physical features can be used to distinguish genus *Lophyra* from other genera of tribe Cicindelini (Fig. 6).

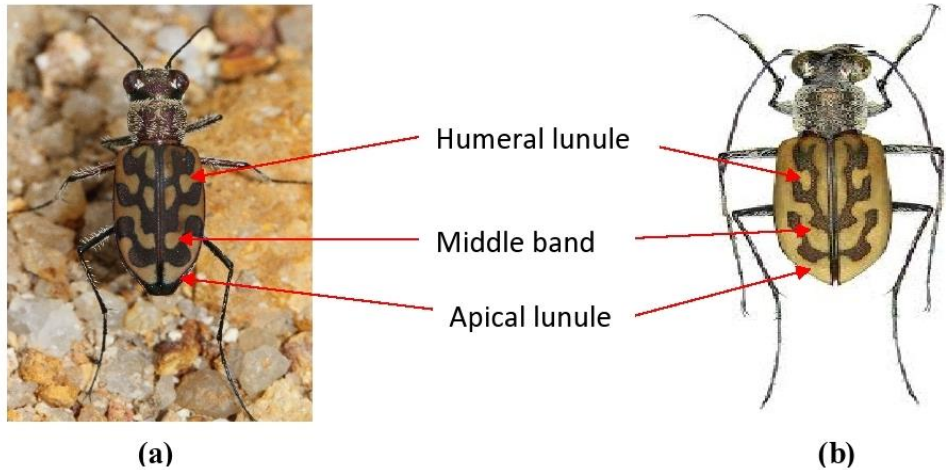

**(a)** **(b)**

**Fig. 6.** (a) *Lophyra catena*; (b) *Lophyra cancellata*

## IV.    Genus *Hypaetha*

Images of two species *Hypaetha biramosa, Hypaetha quadrilineata* were included into this genus. More than 90% of the elytral surface of *Hypaetha biramosa* is covered with dark brown/black maculae medially and a yellowish-white continuous band extends from humeral angle to the apex. This band invaginates around half of the length of the elytra to separate the dark brown/black maculae into two sections (Fig. 7a).

*H. quadrilineata* maculae exist as two longitudinal yellow-white bands on an elytron (Fig. 7b). The abdomen of *H. quadrilineata* is setose laterally, while it is glabrous in *H. biramosa* [22].

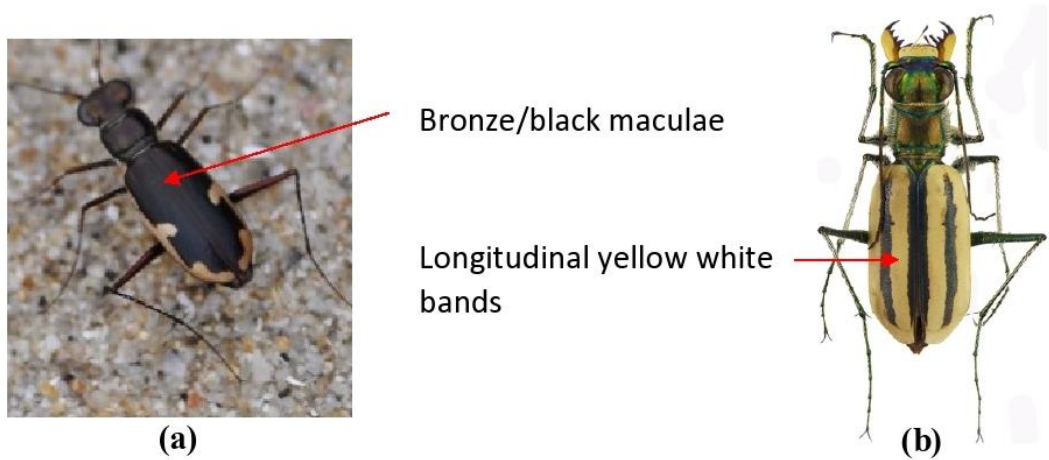

**(a)** **(b)**

**Fig. 7.** (a) - *Hypaetha biramosa; (b) - Hypaetha quadrilineata*

## V.  Genus *Myriochila*

*Myriochila distinguenda* and *Myriochila* (*Monelica*) *fastidiosa* represent the genus *Myriochila* of Sri Lanka. Both species are characterized by the standard pattern of maculae on the elytra consisting of humeral lunule, middle band and apical lunule. In *Myriochila distinguenda* the basal portion of the elytral humeral lunule is separated while in *Myriochila* (*Monelica*) *fastidiosa* it is joined to the apical end of the humeral lunule (Fig. 8).

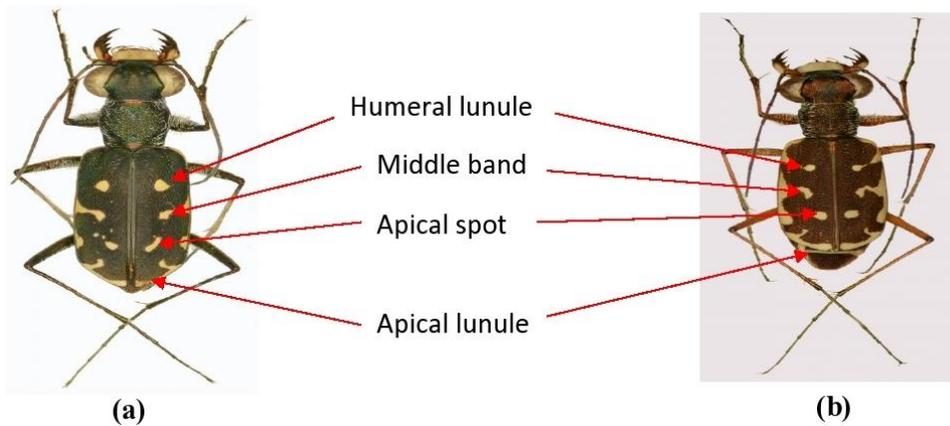

**Fig. 8.** (a) *Myriochila distinguenda;* (b) *Myriochila* (*Monelica*) *fastidiosa*

## VI.  Genus *Cylindera*

The species belonging to this genera are characterized by yellowish-white spots on elytra and do not have humeral, apical lunules or marginal/ middle bands. However, the position, number and shape of the elytral spots in each species of *Cylindera* may vary distinguishing the species from one another (Fig. 9) [3, 14].

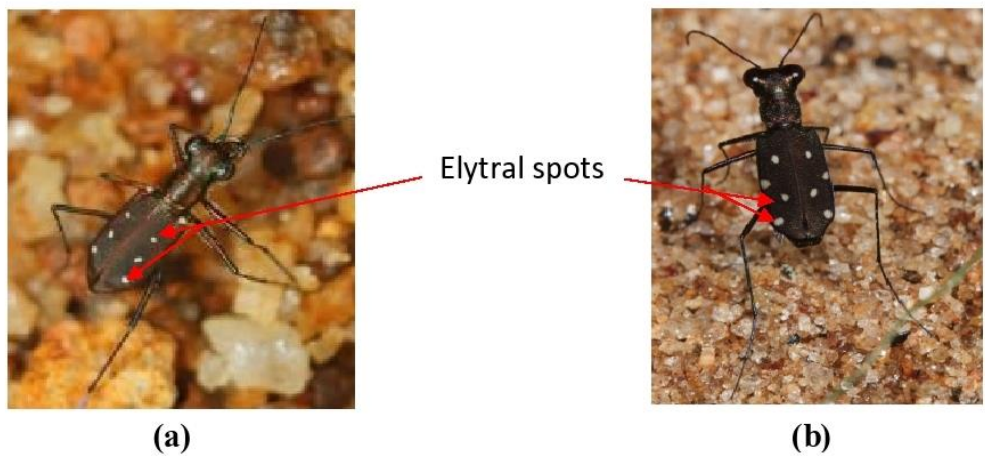

**Fig. 9.** (a) *Cylindera* (*Ifasina*) *waterhousei;* (b)  *Cylindera* (*Ifasina*) *labioaenea*

## 2.1.2 Tribe Collyridini

The main differences between the genera of tribes Collyridini and Cicindelini are found related to the body size, body colour and characteristics of pronotum. In the present study, three genera of tribe Collyridini were considered, *Tricondyla* and *Derocrania* of sub-tribe Tricondylina, and *Neocollyris* of sub-tribe *Collyridina*.

### *I.*   Genus *Derocrania*

Species of this genus are smaller and slender than species of genus *Tricondyla* of the same sub-tribe. Eyes are prominent and pronotum more elongate, slender and narrow at the top. Elytra is elongate and almost widened behind or sometimes in some species very strongly raised behind or almost flat, with very variable sculpture. Legs are long and slender (Fig. 10) [9, 14].

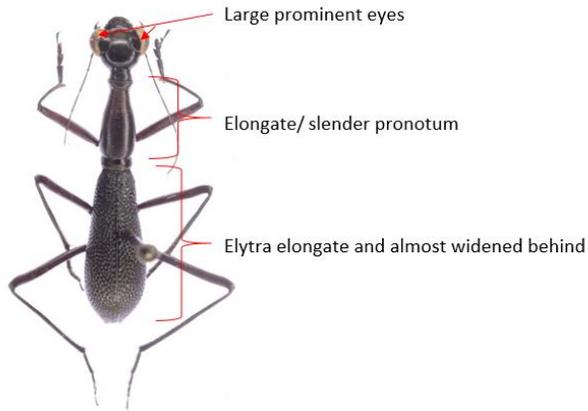

**Fig. 10.** *Derocrania schaumi*

## II.     Genus *Tricondyla*

Species of genus *Tricondyla* are characterized by a large head that is deeply excavated between eyes. Neck behind eyes is parallel-sided. Elytra narrowed in front and dilated and very convex behind. The pronotum is almost parallel sides, broad, constricted in front and behind, sometimes a little convergent but without a collum in front (Fig. 11) [9, 14].

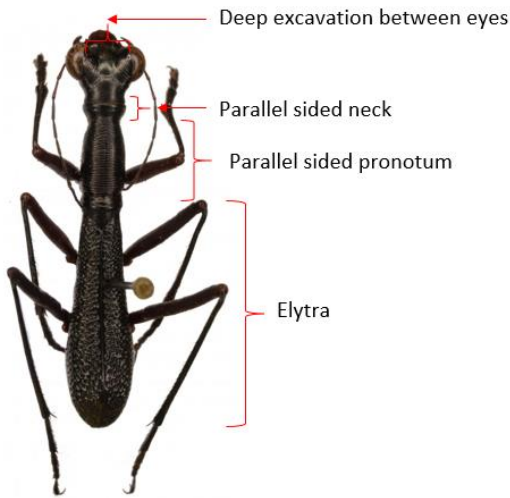

**Fig. 11.** *Tricondyla granulifera*

### III. Genus *Neocollyris*

Species of this genus are small and slender with bright blue elytra that are almost punctured. However, elytra vary in colour, size and sculpture and in some species it is strongly rugose in the middle. Many species of this genus are hard to identify from one another. However, differences are found in the shape of the pronotum which is generally flask-shaped (Fig. 12) [14, 36].

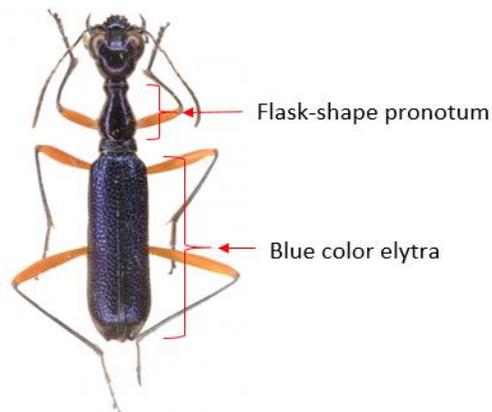

**Fig. 12.** *Neocollyris bonelli*

### 2.2 Image collection

The dataset comprises of images taken during field investigations, from wildlife and nature photographers using different camera types with different image quality, tiger beetle publications and web sites.

However, due to the rarity of certain species and endemicity, only a limited number of images were collected from each genus (Fig. 13).

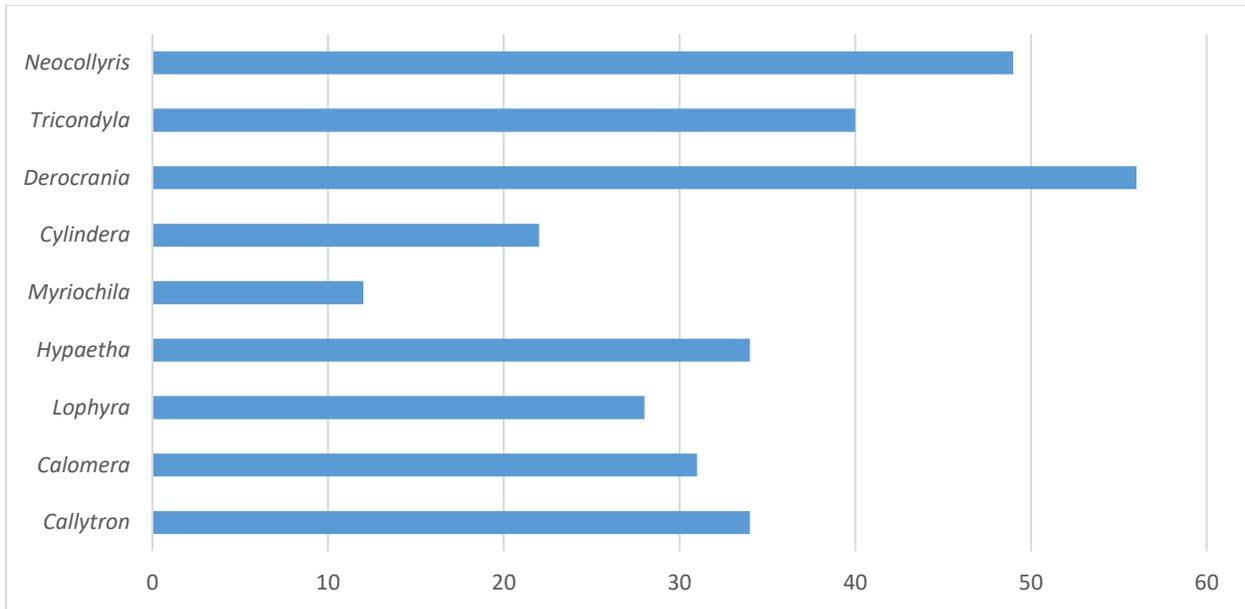

**Fig. 13.** Image samples collected for each genus

## 2.3 Features of the images

## 2. 3.1 Image sources

Images were collected from different sources

- Mobile phone captures
- DSLR camera captures
- Images from websites/blogs (iNaturalist, Shnao, Project Noah, JungleDragon, Thailandwildlife , My Shot Gallery of Bengkulu)

Therefore the images have different quality (Fig. 14), views and angles (Fig. 15).

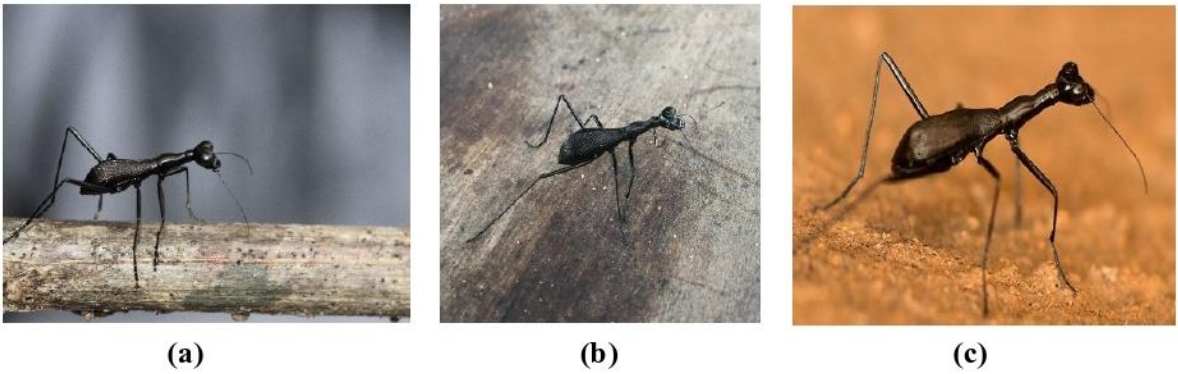

**Fig. 14.** Images obtained from different sources (a) DSLR (macro lens) camera capture; (b) smartphone captures; (c) image taken from a website (genus- *Derocrania)*

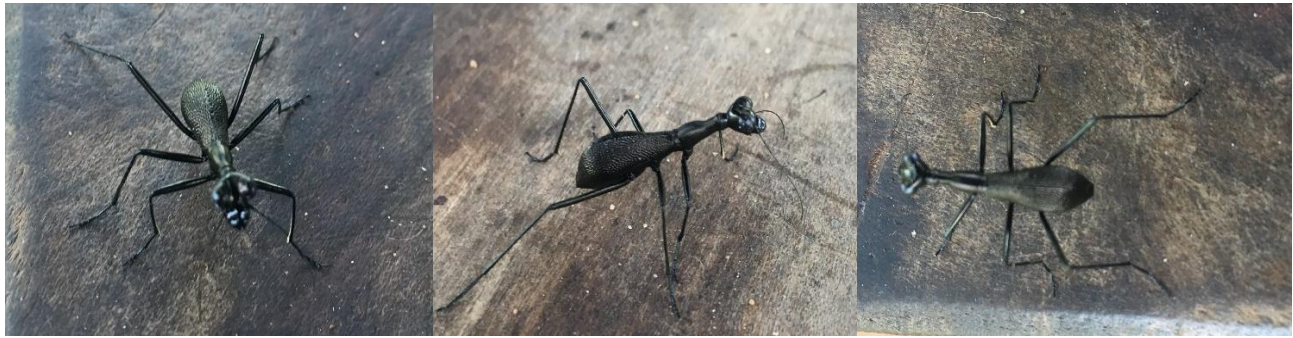

**Fig.15.** Images of species taken in different angles (genus-*Derocrania*)

## 2.3.2 Image background

Furthermore, salient parts in each image have extremely large variations in size (Fig. 16). Subsequently, most of these images are with noisy backgrounds and consistency as most of these images were taken in the actual habitat environment of the species (Fig. 17).

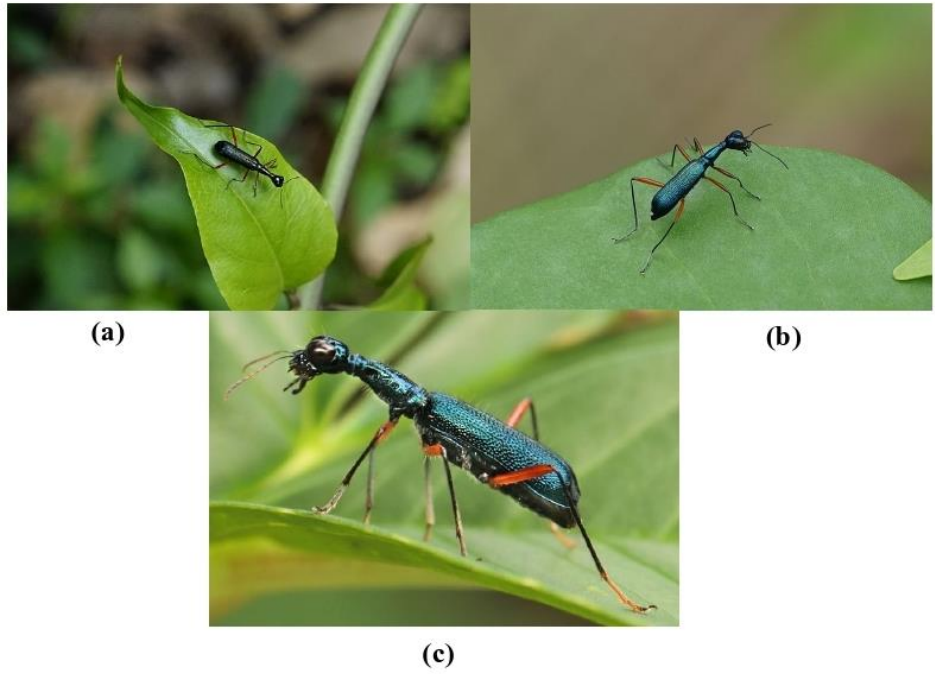

**Fig. 16.** (a) Species occupying very little space; (b) species occupying a part of it; (c) species occupying most of the image (genus- *Neocollyris*)

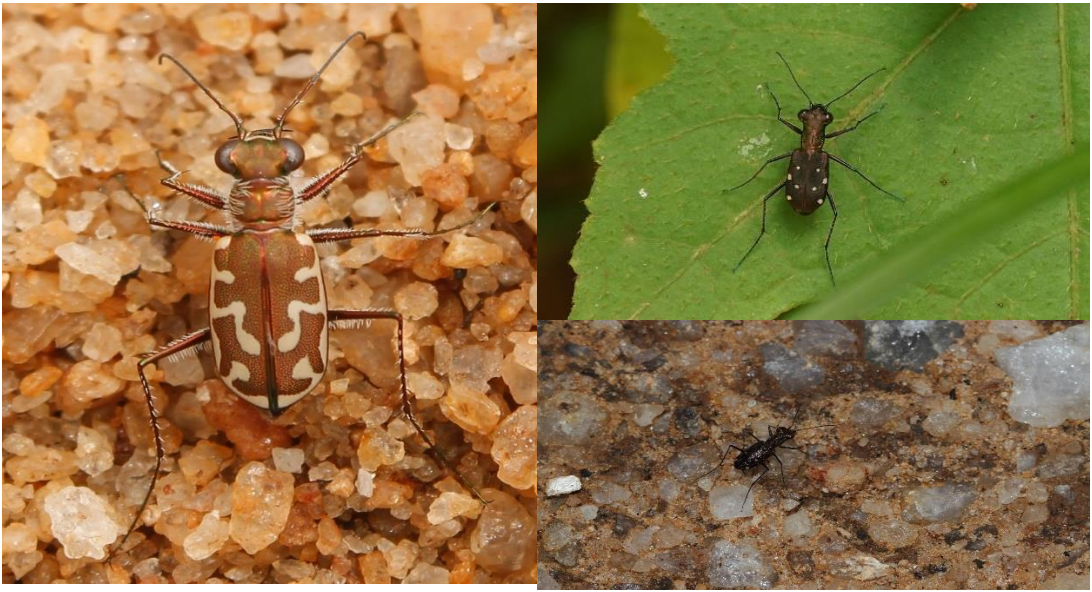

**Fig.17.** Images with background noise

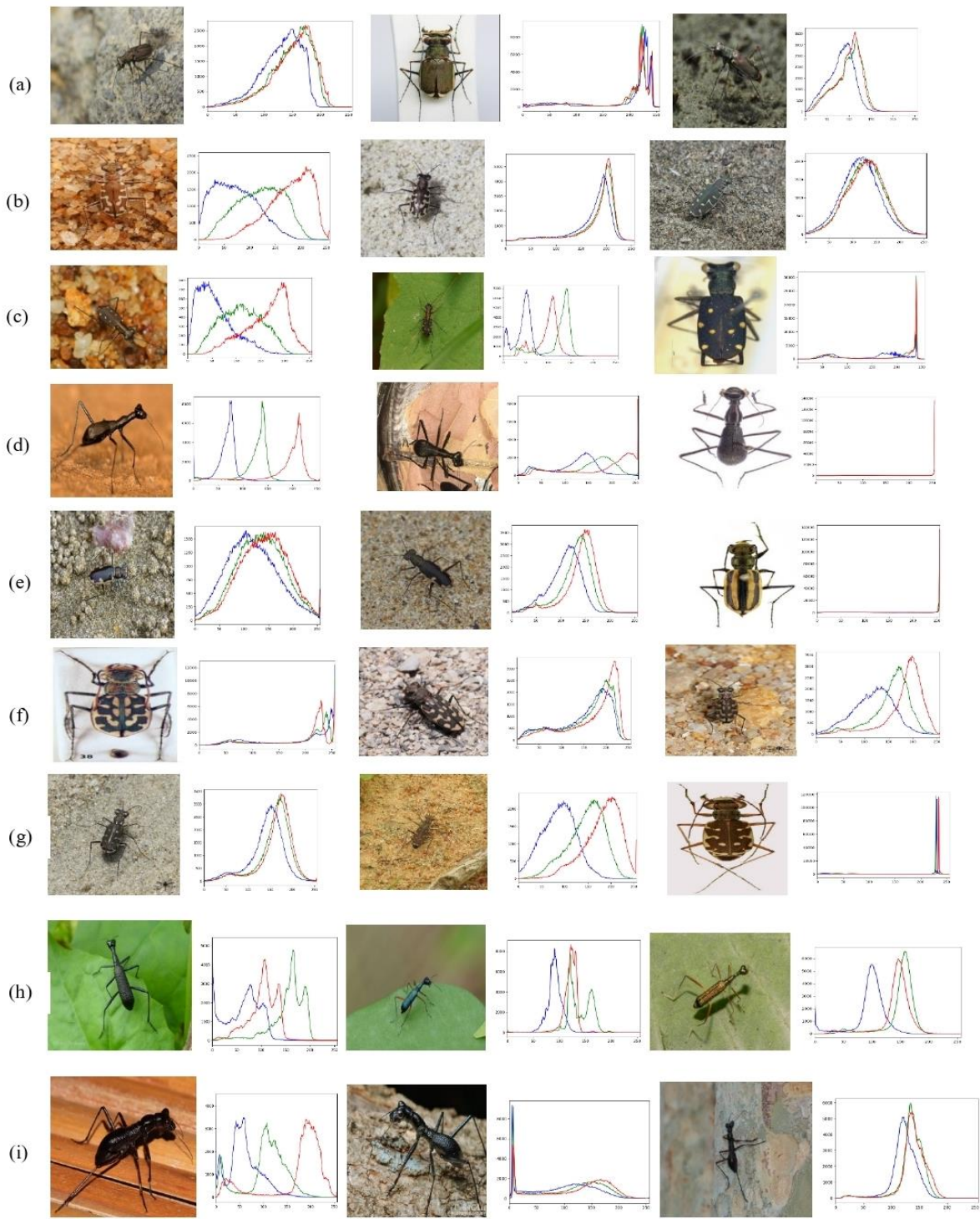

**Fig. 18.** Sample images of each genus and corresponding colored histogram related to the specific image ((a) *Callytron;* (b) *Calomera;* (c) *Cylindera;* (d) *Derocrania;* (e) *Hypaetha;* (f) *Lophyra;* (g) *Myriochila;* (h) *Neocollyris;* (i) *Tricondyla*)

### 2.3.4. Image diversity based on colour saturation

In order to understand certain features of the images and to obtain an overall idea about the colour intensity distribution, combined colour histograms of images for each genus of the dataset were plotted (Fig. 18). Each graph is plotted with pixel values (ranging from 0 to 255) in X-axis and the corresponding number of pixels in each image on Y-axis [16, 39]. The graphs depict variations in colour contrast, brightness, the intensity distribution of images due to background noise of images resulting from their environment.

Limitations in the dataset

- Imbalanced image count - The dataset is highly imbalanced, with certain genera having more images than others.
- Background noise in images.
- Availability of a limited number of images due to endemicity and rareness of species.
- Different quality images.
- Images taken in different angles/ dimensions.

# 3 Classification Algorithms

## 3.1 Classical machine learning classifiers

As the current task is related to supervise learning category where it is required to identify the dependencies between the target prediction output and the input features, and as the image dataset comprises of nine genera of subfamily Cicindelinae, multi-class classification should be done. Before passing images to multi-class classifications, feature extraction needs to be done as preprocessing steps to perform the classification task more effectively. When deciding about the features that could quantify these nine genera of tiger beetles, variations in texture, color patterns (Fig. 18) and shape were considered as global feature vectors, which describe the image as a whole to generalize the entire object. Texture defines the consistency of patterns and colours in an object/image. Since tiger beetles can be identified to genus level from combinations of patterns and colour variations of elytra, texture can be selected as a promising feature to extract unique features of each class (which helps to identify class separately) in the present dataset. However, when considering the shape as a single vector for feature extraction, it is less likely to produce good results since classes of the current dataset have many attributes in common due to inter-class similarities (Fig. 1). Therefore we combined both colour and shape as a single feature descriptor in order to describe the image more effectively. Texture based feature extraction can be done based on different concepts such as Ant Lion Optimizer(ALO)[2, 51], Grey Level Co-occurrence Matrix (GLCM)[15]. For the current process texture was quantified using haralick texture features[21] which are calculated from Grey Level Co-occurrence Matrix, (GLCM), a matrix that counts the co-occurrence of neighbouring grey levels in an image. Further, colour variations and shape were quantified using colour histograms[4] and Hu Moment[23] respectively. After the feature selection process data were passed through eight supervised, multi-class classification algorithms [31] (Fig. 19).

I. Logistic Regression - As the current scenario is a multiclass classification problem a multinomial logistic regression approach was implemented by using Broyden–Fletcher–Goldfarb–Shanno (bfgs) algorithm as the optimization algorithm.

II. Linear Discriminant Analysis (LDA) - The objective of LDA is to project the dataset onto a lower-dimensional space with good class- separability to avoid overfitting and also reduce computational costs.

III. KNeighborsClassifier - It is important to find the best K (neighbour) value to obtain the optimum accuracy from KNN for the dataset. Therefore, to get the best possible fit for the images set K=1 was selected.

IV. DecisionTree Classifier - Classification tree was selected for the current problem. This algorithm performs variable screening/feature selection implicitly which is an advantage for feature extraction. Tree-based learning algorithms are considered as the best and frequently used supervised learning methods as they assist predictive models with high accuracy, stability and affluence of interpretation. Unlike linear models, these algorithms are able to map non-linear relationships well.

V. RandomForest Classifier - RandomForest is a classification algorithm that evolves from decision trees. It consists of a collection of a large number of individual decision trees. For the current problem, 25 decision trees were used in the forest. To classify a new instance, each decision tree provides a classification for input data and this classification algorithm collects the classifications and chooses the most voted prediction as the result [33].

VI.    GaussianNB - When Naïve Bayes is extended to real-valued attributes by assuming Gaussian distribution it's called Gaussian Naive Bayes.

VII.   SVM - Support Vector Machine algorithms are based on the idea of decision planes (hyperplane) which outline decision boundaries which help to classify the data points. SVM algorithms use a set of mathematical functions that are defined as the kernel. The purpose of the kernel is to take data as input and transform it into the required form. Different SVM algorithms use different types of kernel functions. RBF kernel function was used for this scenario [13].

VIII.  ExtraTreesClassifier - An ensemble learning technique. This is very similar to random forest classifier and differs only from the manner of construction of the decision trees[47]. The number of decision trees used in the classifier was 10.

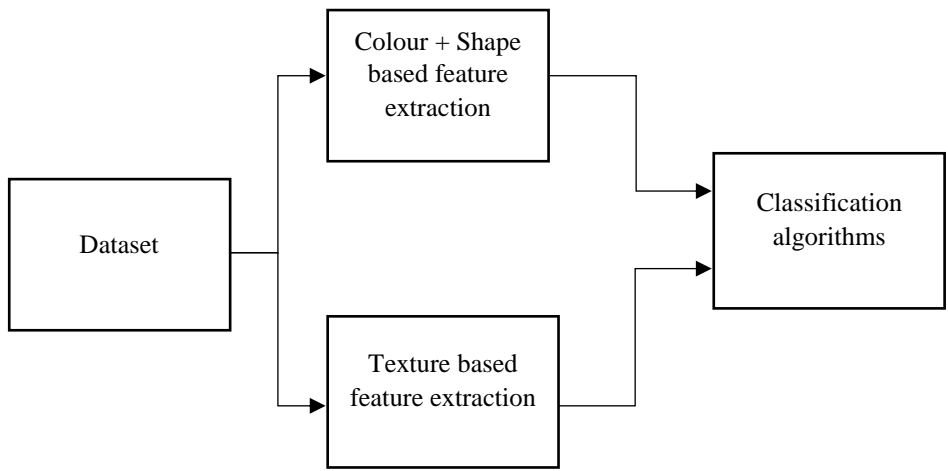

**Fig. 19.** Feature extraction and classification.

Above mentioned traditional machine learning methodologies (classification algorithms) use trivial structures to handle limited data and computing units. When the target objects have rich meanings with high background noise and inconsistency the performance and generalization ability of the above models are insufficient[13]. Therefore it is important to move to deep learning methodologies which can attain higher accuracy. The ultimate goal of deep learning is to represent an image in a hierarchical manner, increasing complexity per each layer. Where natural word objects are also a composition of many combinational units which increase the diversity of resulting structure (Fig. 20).

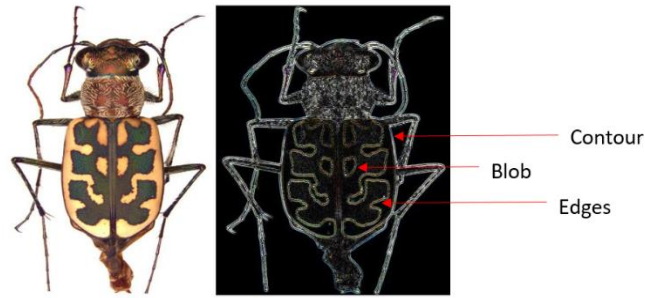

**Fig. 20.** Entomological view against computer vision based descriptors. Genus: *Lophyra*

## 3.2. Deep learning techniques

The pre-processing required for deep learning models are much lower when compared with traditional classification algorithms since these approaches work by extracting features from images and eliminate the need of manual feature extraction. At the same time arrangement of these models play a major role in designing and creating new architectures for the improvement in performance. Therefore, model evaluations instigated from simple structures to complex form, since convolutional neural networks arguably considered a black box evaluation [6]. Based on strategy these models were evaluated on 3 stages.

- CNN models without using pre-trained weights.
- CNN models fine-tuned with transfer learning approach (using pre-trained weights).
- Extract features from pre-trained CNN models, and classify using SVM.

**I. CNN models without using pre-trained weights.**

- **CNN model with 4 convolutional layers and 2 fully connected layers**

  Convolutional neural network model comprises of 6 hidden layers including 4 convolutional layers and 2 fully connected layers. First and second convolutional layers of the model contain 32 kernels of size 5x5, while last or the third convolutional Layer contains 64 kernels of size 3x3 and the final convolutional layer contain 96 kernels of 3x3.

- **AlexNet model**

This is a well-known CNN model which follows a standard neural network architecture of stacked and connected layers. It comprises of eight layers that need to be trained, five convolutional layers followed by three fully connected layers, as well as max-pooling layer[29, 45, 53].

- **SqueezeNet model**

Through channel projection bottleneck (squeeze layers), this model features a great reduction in parameter space and computational complexity. Further, similar to residual networks, the model uses indemnity-mapping shortcut connections which allow for stable training of deep network models. This model is comprised of "fire modules", where the input map is first fed through a bottlenecking channel-projection layer and then divided into two-channel sets. The first channel set is expanded through a $3 \times 3$ convolution and the second one through channel projection. The final convolution map is globally average-pooled into a 512-vector and then fed to a fully-connected layer with 2048 units [17].

**II. CNN models that are fine-tuned with transfer learning strategy (using pre-trained weights)**

Transfer learning has become a conventional procedure (meaning that a classifier is already trained on a large-scale dataset like ImageNet dataset before the actual training begins). Here the classifier will only be fine-tuned to the specific classification problem by training a small number of high-level network layers proportional to the amount of available problem-specific training data[37]. Since the model has already learned certain features from a large dataset this method is more suitable for these kind datasets which are with a limited number of data. Transfer learning (TL), is commonly used in the computer vision area which allows building more precise models efficiently [24, 42, 44].

Therefore as the initial phase, it is required to get pre-trained models with weights loaded. Five deep learning networks were selected as target models.

- AlexNet
- InceptionV3
- ResNet
- SqueezeNet
- VGG16

For pre-trained weights, we used weights gained by training above models on top of ImageNet dataset to extract general classification-supporting features like curves and edges from several front layers.

As the proposed architecture we removed final fully connected layer from each model and then used the remaining portion of the model as a feature extractor for the current dataset. These extracted features are called "Bottleneck features".

Then the fully connected layers were developed according to target dataset below the extracted bottleneck features in order to get the classes as outputs for the problem. Finally, results were classified using softmax (multi-nomial logistic regression) classifier.

### III.    Extract features from pre-trained CNN models, and classify using SVM

Image features from different CNN architectures pre-trained on the ImageNet data set submitted to a linear support vector machine classifier, which is trained on the target problem. For this strategy, we selected the same Convolutional Neural Network models which were used for earlier evaluation (AlexNet, InceptionV3, ResNet, SqueezeNet and VGG16) as target pre-trained deep learning models to extract features. The main reason to select support vector machine algorithm as the classifier was SVMs are more appropriate for small data sets where the number of dimensions are greater than the number of examples [49]. Further SVMs are memory-efficient since they only use a subset of training points or support vectors and they generalize well to high dimensional spaces as well.

# 4. Results and Discussion

## 4.1 Experimental setup

The original images set was divided into a training and test set where around 75% of the total images in each genus were put into training set while the rest (25%) were placed into a test set.

As deep artificial neural networks require a large corpus of training data to learn effectively and to avoid over-fitting. To increase the image quantity, dataset expanded artificially by image augmentation [41, 43] using different ways of processing or combination of multiple processing, such as random rotation, shifts, shear and flips, etc. For this, we used Augmentor [8], a Python package designed to aid the augmentation and artificial generation of image data for machine learning tasks (Fig. 21).

The augmentation procedures used to increase the amount of available training data are listed below.

- Horizontal flipping
- Vertical flipping
- Zooming - Zoom in to an image at a random location within the image, while maintaining its size. The amount by which the image is zoomed is a randomly chosen value.
- Rotate - $90^0$, $45^0$, $180^0$

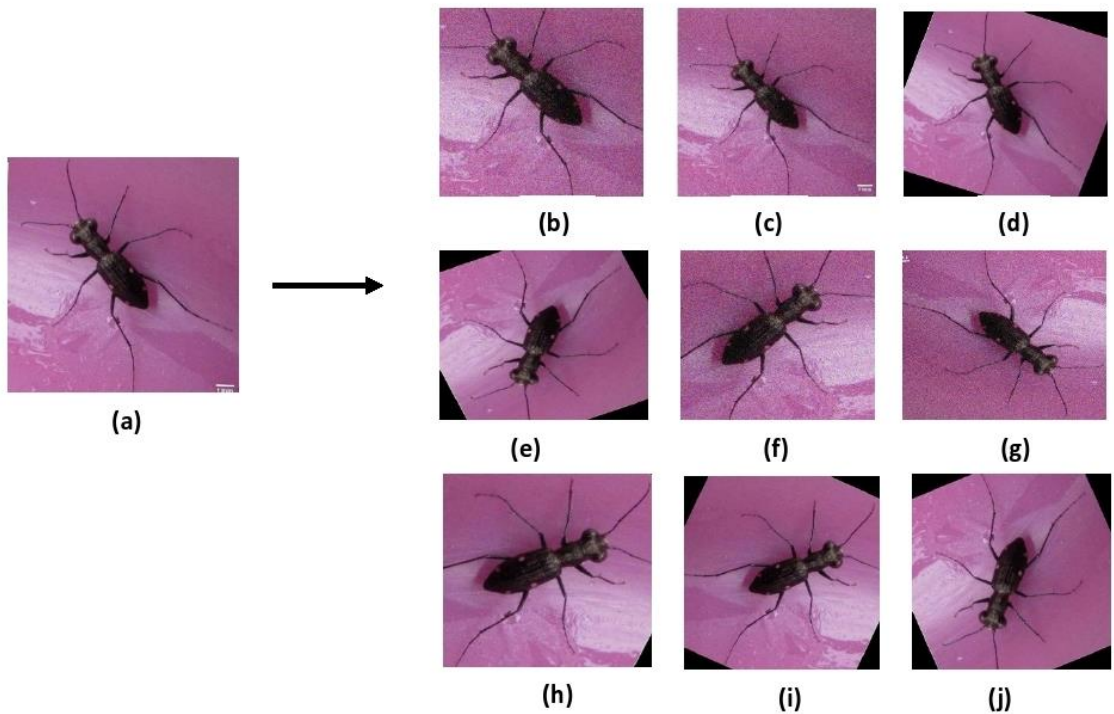

**Fig. 21.** Image Augmentation (a) Original image. (b)Zoomed-in image. (c) Zoomed-out image. (d) $45^0$ rotated image to right (e) $90^0$ rotated image to left (f) Vertically flipped and zoomed-in image (g) $180^0$ rotated image to right (h) $90^0$ rotated image to right (i) $90^0$ rotated and zoomed-out image (j) $90^0$ rotated image to left

After data augmentation process the training image quantity increased up to 100 images per each genus and test image quantity to 35 for each genus, which is still a low image quantity for a deep convolutional neural network. However, increasing image quantity to thousands using image augmentation is also unreasonable since it will again subject the model to over-fit.

Finally, two approaches were initiated, one with original, imbalanced data (Approach I) and other with augmented, balanced data (Approach II) (Table 2).

**Table 2**. Separation of images to train and test set and creation of 2 datasets as Approach I and Approach II

| Tribe | Genus | Approach I | | Approach II | |
|---|---|---|---|---|---|
| | | Original training image quantity | Original testing image quantity | Training image quantity (after augmentation) | Testing image quantity after augmentation) |
| Cicindelini | *Callytron* | 29 | 6 | 100 | 35 |
| | *Calomera* | 39 | 5 | 100 | 35 |
| | *Lophyra* | 33 | 4 | 100 | 35 |
| | *Hypaetha* | 37 | 4 | 100 | 35 |
| | *Myriochila* | 37 | 4 | 100 | 35 |
| | *Cylindera* | 20 | 7 | 100 | 35 |
| Collyridini | *Derocrania* | 37 | 21 | 100 | 35 |
| | *Tricondyla* | 35 | 7 | 100 | 35 |
| | *Neocollyris* | 44 | 11 | 100 | 35 |

In order to perform a benchmark on the dataset, the previously mentioned classification algorithms were trained on the datasets of Approach I (original, imbalanced data) and Approach II (augmented, balanced dataset) while deep learning models were trained on only the augmented, balanced dataset.

## 4.2 Test Results and Discussion

**Table 3.** Test accuracies of classification algorithms

| Feature vector | Colour+Shape | | | | | | | | Texture | | | | | | | |
|---|---|---|---|---|---|---|---|---|---|---|---|---|---|---|---|---|
| Classification algorithms | | | | | | | | | | | | | | | | |
| | LR | LD | KNN | DT | RF | GNB | SVM | ETC | LR | LD | KNN | DT | RF | GNB | SVM | ETC |
| Original imbalanced images set (Test accuracy %) | 39.13 | 8.70 | 37.68 | 26.08 | 42.02 | 31.88 | 15.94 | 39.13 | 44.94 | 49.43 | 30.34 | 42.69 | 44.94 | 47.19 | 31.46 | 43.82 |
| Augmented balanced images set (Test accuracy %) | 36.82 | 13.96 | 33.97 | 26.34 | 38.09 | 23.17 | 11.11 | 37.77 | 26.35 | 39.37 | 26.67 | 31.11 | 40.63 | 27.61 | 19.68 | 37.46 |

According to Table 3, classification algorithms using texture-based feature extraction is more suitable for the classification of the current dataset than classification algorithms using colour and shape. The highest test accuracy of 49.43% has gained from linear discriminant analysis algorithm based on texture-based feature extraction. Further higher validation accuracies have gained from all the classification algorithms which were trained on original imbalanced images set than augmented balanced images set. Therefore, it is able to state that the orientation of the images supports the machine learning models for classifications and orientation is a key factor for image classification using a machine learning approach.

**Table 4.** Test accuracies of Convolutional Neural Network models

| Deep learning model | Val top 1 accuracy | Val top 2 accuracy | Val top 3 accuracy |
|---|---|---|---|
| AlexNet | 34.28% | 39.68% | 57.46% |
| CNN with 4 convolutional layers and 2 fully connected layers | 42.1% | 50.25% | 61.28% |
| SqueezeNet | 63.81% | 69.21% | 79.68% |

When considering deep learning models which were trained without using pre-trained weights on augmented balanced tiger beetle images set, squeezeNet gave the highest validation accuracy of 63.81% (Table 4).

**Table 5**. Model complexity with validation accuracies of different Convolutional Neural Network models

| Deep learning model | Number of layers | Number of trainable parameters | Val top 1 accuracy |
|---|---|---|---|
| AlexNet | 8 | 28,067,625 | 34.28% |
| CNN with 4 convolutional layers and 2 fully connected layers | 6 | 889,673 | 42.1% |
| SqueezeNet | 2 | 543,387 | 63.81% |

According to Table 5, the validation accuracies have increased when the complexity of the model decreases. Therefore, it can be assumed that model complexity is not always the solution for better accuracy. At the same time [25] elaborate that CNN models with fewer parameters have several advantages such as require less communication across servers during distributed training and more feasible to deploy on FPGAs and other hardware with limited memory.

**Table 6**. Test accuracies of pre-trained CNN+SVM models

| Pre-trained CNN+SVM | Val top 1 accuracy |
|---|---|
| SqueezeNet | 60.20% |
| InceptionV3 | 40.25% |
| ResNet | 41.27% |
| AlexNet | 55.24% |
| VGG16 | 60% |

`

**Table 7**. Test accuracies of Transfer learning models using Softmax classifier

| Transfer learning model | Val top 1 accuracy |
|---|---|
| SqueezeNet | 64.44% |
| InceptionV3 | 42.83% |
| ResNet | 69.52% |
| AlexNet | 71.75% |
| VGG16 | 75.87% |

Table 6 and Table 7 shows that models which were trained using transfer learning approach have given more promising accuracies than models that extracted features using pre-trained CNN models and classified using SVM. Therefore, the test accuracies of pre-trained CNN-Softmax are higher than pre-trained CNN-SVM when relating to the current dataset.

The highest test accuracy for the dataset was gained from the transfer learning VGG16 model which was 75.87% (Table 7).In this model as an additional modification all layers above fully connected layers were not trained (freeze) thus only fully connected layers were considered as trainable parameters. When using weights of a pre-trained model on top of the current model, the complexity of the CNN model has not become a significant factor for the improvement of validation accuracies (Table 8). From the above approach, even model with least number of trainable parameters has gained the optimum accuracy for highly diverse limited image tiger beetle dataset. The results depict that better accuracies for vision-based datasets with limited data can be gain from machine learning models with fewer trainable parameters.

**Table 8.** Model complexity with validation accuracies of different transfer learning models

| Deep learning model | Number of layers | Number of trainable parameters | Val top 1 accuracy |
|---|---|---|---|
| InceptionV3 | 48 | 889,673 | 42.83% |
| SqueezeNet | 2 | 727,113 | 64.44% |
| ResNet | 50 | 23,451,209 | 69.52% |
| AlexNet | 8 | 16,002,633 | 71.75% |
| VGG16 | 16 | 4,617 | 75.87% |

Validation accuracies gained for tiger beetle image dataset from classification algorithms were significantly low and although the image quantities were increased using image augmentation, there were no significant differences among validation accuracies before and after image augmentation. However, accuracy gain from texture-based feature extraction was bit higher than colour and shape-based feature extraction method. Further, when considering the results gained from different deep learning models it clarifies that higher validation accuracies have gained from transfer learning approaches using softmax classifier. This is mainly because pre-trained models have trained over ImageNet dataset which contains images of tiger beetles as a single collection under beetle hierarchy. Hence the models have learned some features from larger dataset previously it aid to identify features in the new dataset. Therefore, the baseline performance of the model also improves due to knowledge transfer. Here highest accuracy was gained from VGG16 transfer learning model. Due to reasons like inter-class similarity and availability of limited data, machine learning models have gained below 80% validation accuracies, which has plenty of room for further improvement. The above experiments were conducted to provide a benchmark and above validation accuracies depict the diversity of the tiger beetle image dataset.

## 5. Conclusion

The present article reveals a highly precise and diverse dataset created for tiger beetles (Coleoptera, Cicindelinae) of Sri Lanka which contains images of beetles of nine genera of two tribes. Images were tested on different classification algorithms based on different feature extraction techniques (texture, colour, shape), and deep learning models with and without using pre-trained weights. Further, an augmented and balanced dataset was evaluated by extracting features from pre-trained CNN models and classifying them using SVM classifier. The study produced a dataset that was highly specific and challenging for renowned machine learning models and beneficial for automated zoological studies since this is a real-time dataset, consisting of features like high unevenness of data, a limited number of data with huge background noise and having lots of inter-class similarities in images. As benchmark results from above models, optimum test accuracy was obtained from transfer learning models where VGG16 transfer learning model gave the highest test accuracy of 75.87%. Attempting to improve accuracies for these types of datasets will assist to overcome limitations in image processing techniques and expand machine learning knowledge. Further, these attempts will be beneficial in expanding biodiversity monitoring systems on a global scale.

**Acknowledgements.** This work was funded by the National Science Foundation of Sri Lanka [Grant number RG/2017/EB/01].

Uppasala University,Sweden

**Conflict of interest:** The authors declare no conflict of interest.

**Availability of data and material:** Tiger beetle dataset which used for the evaluation can access through following URL (using given username and password)

    **URL:** **https://fos.cmb.ac.lk/opendata/tigerbeetles/**

    **Username: lakmini**

    **Password: VDI7SF9Z**

**Code availability**: Sample machine learning models can access through following URL (using given username and password)

    **URL:** **https://fos.cmb.ac.lk/opendata/tigerbeetles/**

    **Username: lakmini**

    **Password: VDI7SF9Z**

**Ethics approval:** Not applicable

**Consent to participate:** Not applicable

**Consent for publication:** Not applicable